\documentclass[aps,prc,preprint,groupedaddress,nofootinbib,showpacs,preprintnumbers,amsmath,amssymb,superscriptaddress]{revtex4}

\usepackage{graphicx}
\usepackage{dcolumn}
\usepackage{bm}
\usepackage{amsmath}

\begin{document}

\def\R{\right} \def\L{\left} \def\Sp{\quad} \def\Sp2{\qquad}

\preprint{RCNP-Th02020}

\title{Pion-nucleus optical potential valid up to the
DELTA-resonance region}

\author{L. J. Abu-Raddad} 
\email{laith@rcnp.osaka-u.ac.jp}
\affiliation{Research Center for Nuclear Physics, Osaka University, 10-1 Mihogaoka, Ibaraki,
Osaka 567-0047, Japan}  

\date{\today}

\begin{abstract}
We present in this article an optical potential for the $\pi$-nucleus
interaction that can be used in various studies involving
$\pi$-nucleus channels. Based on earlier treatments of the low energy
$\pi$-nucleus optical potential, we have derived a potential
expression applicable from threshold up to the $\Delta$-resonance
region. We extracted the impulse approximation form for this potential
from the $\pi-N$ scattering amplitude and then added to it kinematical
and physical corrections. The kinematic corrections arise from
transforming the impulse approximation expression from the $\pi-N$
center of mass frame to the $\pi$-nucleus center of mass frame, while
the physical corrections arise mostly from the many-body nature of the
$\pi$-nucleus interaction. By taking advantage of the experimental
progress in our knowledge of the $\pi-N$ process, we have updated
earlier treatments with parameters calculated from state-of-the-art
experimental measurements. 
\end{abstract}

\pacs{PACS number(s):~25.80.Dj,13.75.GX,24.10.Ht}

\maketitle
\section{Introduction \label{intro}}

The purpose of this work is to provide 
a $\pi$-nucleus optical potential valid up to the
$\Delta$-resonance region that can be used in various
studies involving the $\pi$-nucleus interaction. Specifically, we
supply a potential that can provide an integral component in the analysis of
several future and past experiments in pion photo- and
electro-production processes from nuclei at Thomas Jefferson National
Accelerator Facility (Jefferson Lab) and
Mainz~\cite{adampro99,krusche98,schmitz96,koch,gothe95}. Indeed, the
personal motivation for this study is a study of the coherent pion
photoproduction from nuclei in the $\Delta$-resonance region~\cite{rad99,thesis}. 
 
The $\pi$-nucleus elastic scattering has enjoyed significant investigation
from a variety of theoretical perspectives~\cite{JohErn}.  Our
approach here is to extend the prominent work of Stricker, McManus and
Carr (SMC)~\cite{SMC1,SMC2,SMC3,SMC4} on the low energy $\pi$-nucleus
optical potential to higher energies so that it covers the $\Delta$-resonance region. Indeed, our treatment here follows closely the SMC
analysis but extends their treatment to higher energies. We do so by
first not using any low energy approximation. Second, we invoke a
fully covariant kinematics including the nucleus recoil. Third, we
update their treatment by using $s$- and $p$-wave parameters calculated
from state-of-the-art experimental measurements, and we keep these
parameters intact by not attempting to change them to fit any specific
data. Finally, we include terms that the SMC group ignored either because of
their small effect at low energy or because of the fitting procedure
which allowed them to absorb theses terms in the refitted parameters.

Bearing these facts in mind, in what follows we outline the formalism
used to derive the optical potential. We begin from where the SMC work
started
by taking the elementary amplitude of the process $\pi N \rightarrow
\pi N$ and using it, along with the impulse approximation, to develop
the amplitude for the $\pi$-nucleus interaction. We arrive then at
what is known as the impulse approximation form of the optical
potential. Such a form however still lacks two classes of corrections:
kinematical and physical ones. The kinematical ones arise from
transforming the $\pi-N$ elementary amplitude from the $\pi-N$ center
of mass (c.m.) frame to the $\pi$-nucleus c.m. system. The physical
corrections however arise from the fact that the impulse approximation
picture does not encompass distinct many-body interactions
that appear only in the $\pi$-nucleus channel.  These effects include
multiple scattering, pion absorption, Pauli blocking, and Coulomb
corrections. They are of second and higher orders in strength compared
to the first-order expression given by the impulse approximation.

The paper has been organized as follows. In Section~\ref{sec:formal},
the optical potential is derived starting from the $\pi-N$ elementary
amplitude while the final form of the potential is presented in
Section~\ref{sec:form}.  Next, a number of applications 
are discussed in Section~\ref{sec:results}, and various
results are compared with experimental data. Finally, conclusions are
drawn in Section~\ref{sec:concl}.

\section{Derivation of the Optical potential}
\label{sec:formal}

\subsection{Pion-nucleon elementary process\label{piN}}

The starting point for our derivation of the optical potential is the $\pi-N$
 scattering amplitude which is given by~\cite{SMC1,SMC2,SMC3,SMC4}
\begin{equation}
f_{\pi N}(\pi N \rightarrow \pi N) = b_0 + b_1 \; {\bf t} \cdot
\mbox{\boldmath $\tau$ \unboldmath} + (c_0 + c_1 \; {\bf t} \cdot
\mbox{\boldmath$\tau$\unboldmath} ) \; {\bf k}.{\bf k}^\prime\;,
\label{pinamp}
\end{equation}
where ${\bf t}$ and \boldmath $\tau$ \unboldmath are the pion and
nucleon isospin operators, ${\bf k}$ and ${\bf k}^\prime$ are the
incoming and outgoing pion momenta, $b_0$ and $b_1$ are the $s$-wave
parameters while $c_0$ and $c_1$ are the $p$-wave parameters. In this expression
the small spin-dependent term has been neglected~\cite{SMC2}.

The $s$- and $p$-wave parameters are determined from the phase shifts of
the interaction according to a formalism sketched in
Ref.~\cite{ndu91}. In the earlier treatments~\cite{SMC1,SMC2,SMC3,SMC4}, these parameters were determined
initially from a phase shift analysis performed by Rowe, Salomon, and Landau~\cite{RSL}, but then were
modified to obtain the best fit for the $\pi$-nucleus
scattering and pionic atom data. Our treatment here
differs in two respects: first, we extract
the parameters from the state-of-the-art experimental measurements and phase
shift analysis of Arndt, Strakovsky, Workman, and
Pavan from the Virginia Tech SAID program~\cite{Arnpin}.
Second, we keep these parameters intact by not attempting to change
them to fit any specific data. In doing so we have maintained the
theoretical basis for the optical potential unblemished. This is
particularly important in this work as these parameters dominate the optical
potential in the $\Delta$-resonance region.

\begin{figure}
\includegraphics[totalheight=5.5in,angle=-90]{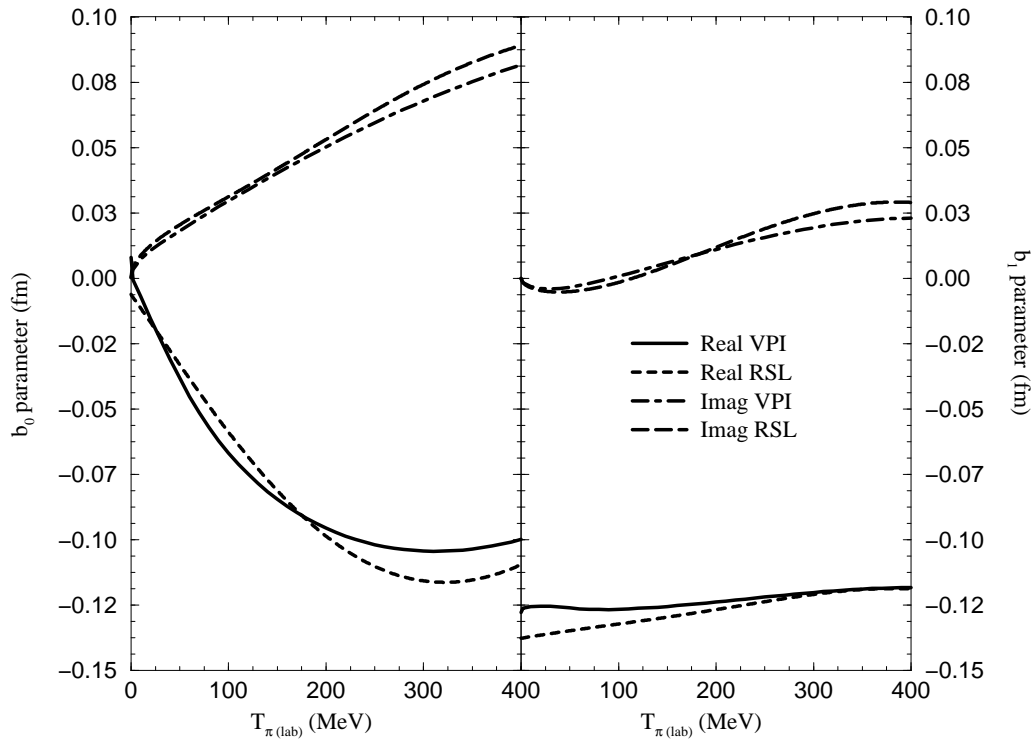}
 \caption{\label{fig1}The $b_0$ (left panel) and $b_1$ (right panel) $s$-wave
elementary amplitude parameters as functions of the pion kinetic
energy in the laboratory system $T_{\pi\;(\text{lab})}$. The figure draws a
comparison between parameters extracted from the state-of-the-art
Arndt {\it et al} (VPI) experimental measurements~\cite{Arnpin} and
those extracted from the Rowe {\it et al} (RSL) phase-shift
analysis~\cite{RSL} more than twenty years earlier.}
\end{figure}

\begin{figure}
\includegraphics[totalheight=5.5in,angle=-90]{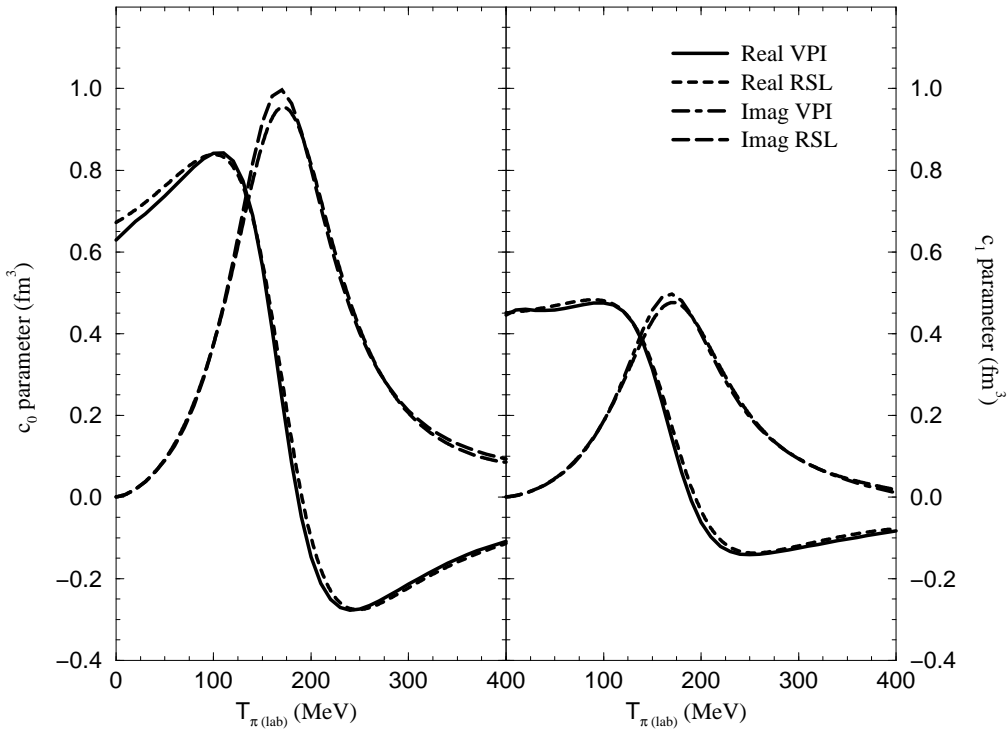}
\caption{\label{fig2} The $c_0$ (left panel) and $c_1$ (right panel) $p$-wave
elementary amplitude parameters as functions of the pion kinetic
energy in the laboratory system $T_{\pi\;(\text{lab})}$. The figure draws a
comparison between parameters extracted from the state-of-the-art
Arndt {\it et al} (VPI) experimental measurements~\cite{Arnpin} and
those extracted from the Rowe {\it et al} (RSL) phase-shift
analysis~\cite{RSL} more than twenty years earlier.}
\end{figure}

Having extracted the $s$- and $p$-wave parameters from the phase-shift
analysis of Arndt {\it et al} (VPI), we compare them with those
extracted using the Rowe {\it et al} (RSL)
phase-shifts. Figures~\ref{fig1} and~\ref{fig2} show the $b_0$, $b_1$, $c_0$, and $c_1$
parameters as functions of the pion kinetic energy in the laboratory
system. It is evident that the VPI and RSL results are
comparable and that there are only small differences. This is indeed
notable in light of the limited experimental measurements at pion
momenta higher than 250~MeV at the time Rowe {\it et al} published their
results.

\subsection{Kinematic corrections\label{kinc}}
	
After adopting the $\pi-N$ amplitude of Eq.~(\ref{pinamp}) in the
$\pi-N$ c.m. frame, the next
step in the derivation is to transform the amplitude to the
$\pi$-nucleus c.m. system. This is done using the relativistic
potential theory of Kerman, McManus, and Thaler~\cite{KeMcTh59} which
establishes a relationship between the $\pi-N$ transition matrix ($t$) in
the $\pi-N$ c.m. frame and the $\pi-N$ amplitude in the
$\pi$-nucleus c.m. frame according to~\cite{heller}:
\begin{equation}
<k^\prime, p^\prime | t | k, p > \; = \; (2 \pi)^3 \delta(\vec
{k^\prime} + \vec{p^\prime} - \vec k - \vec p)\: {\cal{K}} \: f_{\pi N}
({k^\prime}_{cm}, k_{cm}),
\end{equation}
where $k$ and $p$ ($k^\prime$ and $p^\prime$) are the initial (final)
pion and nucleon momenta in the $\pi$-nucleus c.m frame while $k_{cm}$ and
${k^\prime}_{cm}$ are the initial and final pion momenta in the
$\pi$-N c.m. system. $\cal{K}$ is some involved kinematic factor~\cite{SMC2}.

Next, we express the arguments of $f_{\pi N}
({k^\prime}_{cm}, k_{cm})$ in terms of the appropriate kinematical
quantities in the $\pi$-nucleus c.m. frame. This is done using what is
referred to as the ``angle transformation'' which relates the
${k^\prime}_{cm}$ and $k_{cm}$ to the corresponding quantities in the
$\pi$-nucleus c.m. system through a Lorentz
transformation~\cite{SMC2}. As a consequence of this boost, a term
proportional to the dot product of nucleon momenta $\vec{p}\cdot {\vec{p^\prime}}$,
that is proportional to the kinetic energy density in the nucleus, is
encountered. Using Thomas-Fermi approximation, it leads to the optical
potential term $\widetilde{K}(r) \sim c_0
\rho^{5/3}(r)$, where $\rho(r)$ is the nuclear matter density. This
term, although derived by Stricker~\cite{SMC2}, was ignored in the
SMC potential largely because its
effect is absorbed in the parameter fitting. Since we have elected to preserve the
physical basis of our parameters, this term must be included.

\subsection{Multiple scattering corrections\label{muls}}

\begin{figure}
\includegraphics[totalheight=5.5in,angle=-90]{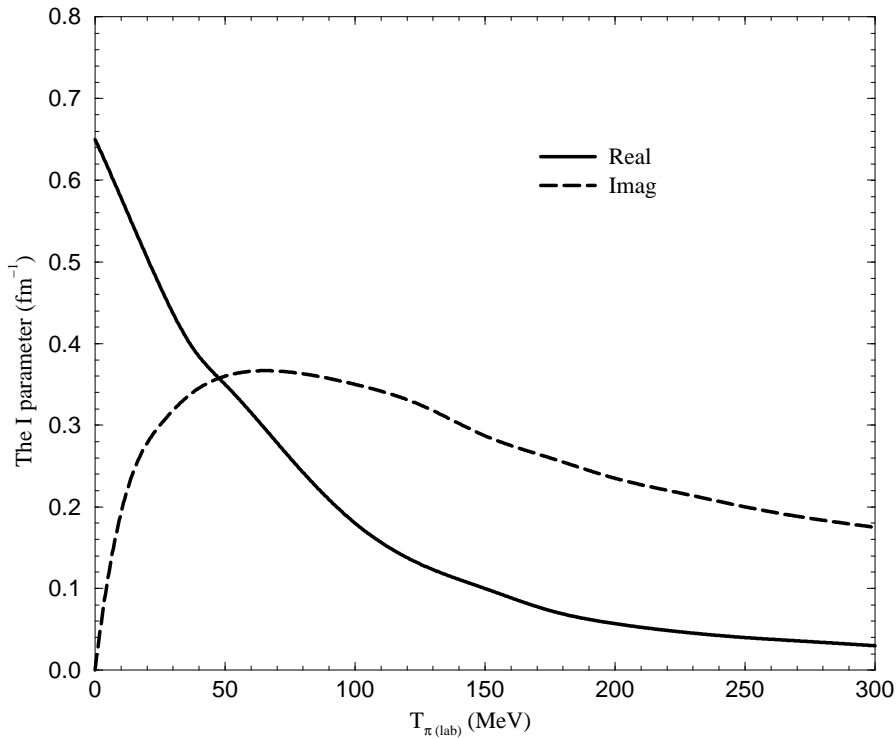}
\caption{\label{I} The $I$ parameter as a function of the pion kinetic
energy in the laboratory system $T_{\pi\;(\text{lab})}$. This quantity is necessary in the
incorporation of multiple scattering corrections to the $b_0$
($s$-wave) parameter.}
\end{figure}

By invoking the impulse approximation, the resultant form for the
amplitude is then sandwiched between bound-nucleon states and the
expression is summed over all occupied states of the nucleus. Hence,
one obtains the $\pi$-nucleus interaction amplitude in momentum
space. Taking the Fourier transform, we obtain the impulse approximation expression for
the optical potential. This form still lacks physical corrections
arising from many-body processes, which alter the
scattering amplitude parameters such as $b_0$ and $c_0$ and add new terms
to the optical potential. Thus, we incorporate the second order
multiple scattering 
corrections to the small (nearly zero) $s$-wave terms that play an
important role only at low energies as 
the $p$-wave effect is still small in this energy regime. This correction
was first calculated by Ericson and Ericson~\cite{erer66}. Here we
adopt the formalism given by Krell and Ericson~\cite{ke69} which
yields a term in the potential proportional to the nuclear density. Effectively, this
correction shifts the $b_0$ parameter by a term proportional to $(b^2_0 + 2 b^2_1)\; I$,
where $I$ is the so-called $1/r_{correlation}$ function
calculated by Stricker~\cite{SMC2} and is shown in Figure~\ref{I}. At
very low energy ($\leq 10$~MeV) a good approximation for this function is $I
= \frac{3 k_F}{2 \pi}$ where  $k_F$ is the nucleon Fermi momentum in
the free gas model. In the SMC
work~\cite{SMC1,SMC2,SMC3,SMC4}, this constant value was used for $I$. It is
clear in Figure~\ref{I} that $I$ falls rapidly and so this
assumption is, in principle, not justified. Nonetheless, since they
fitted their parameters, the effect of this correction is buried in
the refitted value for $b_0$. We include this term in our study
with its exact behavior as a function of energy.

The $p$-wave terms are the crucial ones in the $\Delta$-resonance
region and thus 
we include higher order corrections by summing the multiple scattering
series to all orders. This was first done by Ericson and
Ericson~\cite{erer66} who concluded that this summation introduces a
term of the form $\mbox{\boldmath ${\nabla}$ \unboldmath} \cdot
{\cal Q}(r) \mbox{\boldmath ${\nabla}$ \unboldmath}$. This non-local form
became to be known in the literature as the Ericson-Ericson
effect which is analogous to the Lorentz-Lorenz effect
in electrodynamics~\cite{jackson62}.  The exact form of the function
${\cal Q}(r)$ is a matter of dispute due to the method in which the multiple
scattering series is summed and due to the nature of the assumed short
range correlations between nucleons. Here we follow the SMC
methodology~\cite{SMC1,SMC2,SMC3,SMC4} of summing the series partially to all
orders to obtain a term of the form  
\begin{eqnarray}
  \frac{ L(r) }{1 + \frac {4 \pi}{3} L(r)} + p_1 x_1\:
   \acute{c}\: \rho(r)\;,
\label{eee}
\end{eqnarray}
where $ L(r)$ includes contributions from the $c_0$ ($p$-wave) and
absorption terms to be discussed in Section~\ref{abso}. The kinematic
factors $p_1$ and $x_1$ are defined below. This term still lacks the inclusion of short range
correlations. Various studies have attempted to calculate these
contributions under different assumptions~\cite{BB75,OW79}. They
concluded that such correlations reduce the strength of this form
through a parameter $\lambda$ according to
\begin{eqnarray}
   \frac{ L(r) }{1 + \frac {4 \pi}{3}  \lambda L(r)} + p_1 x_1
   \: \acute{c} \: \rho(r)\;.
\end{eqnarray}
The value of $\lambda$ is not precisely established. Nonetheless, there is an agreement that it is
greater than one due to the finite range of the pion-nucleon
interaction. Baym and Brown predicts a value of 1.6 or higher while
Oset and Weise estimate it in the range of $1.2-1.6$. Values in the range of $1.2-1.6$ are
common in the literature~\cite{SMC1,SMC2,SMC3,SMC4}.

\begin{figure}
\includegraphics[totalheight=5.5in,angle=-90]{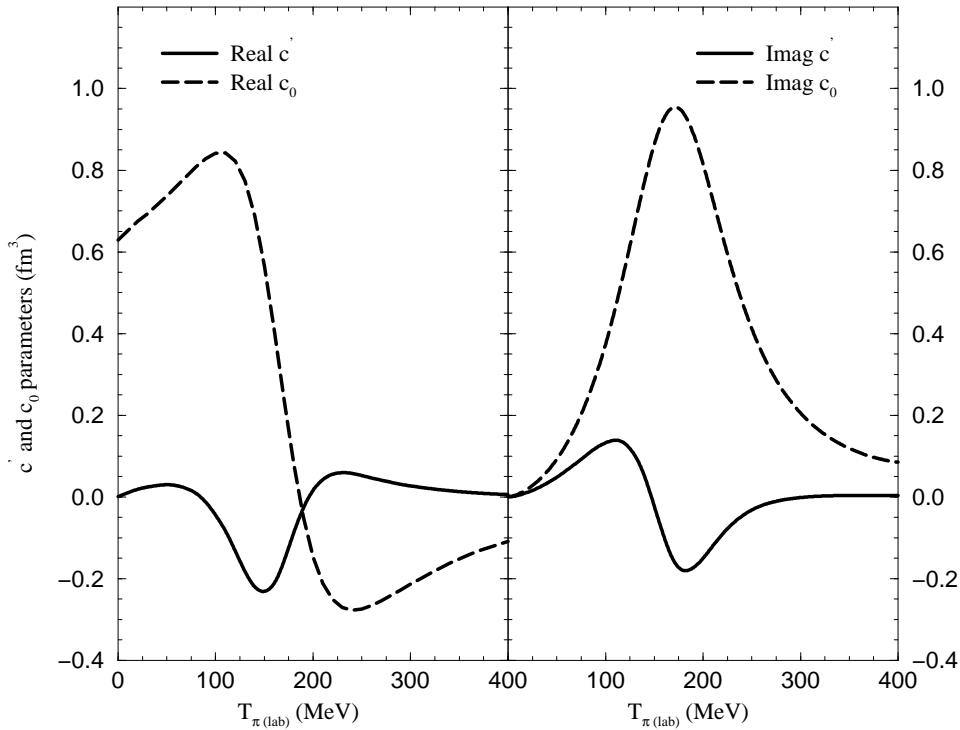}
\caption{\label{ctil} 
The real parts (left panel) and the imaginary parts (right panel) of the $\acute{c}$
and $c_0$ parameters as functions of the pion kinetic
energy in the laboratory system $T_{\pi\;(\text{lab})}$. While small at low and high energies,
the $\acute{c}$ parameter is considerable in the $\Delta$-resonance region.  
}
\end{figure}

The second term $p_1 x_1 \:\acute{c} \:\rho(r)$ in Eq.~\ref{eee} appears
naturally in summing the multiple scattering series~\cite{SMC2} but
was not included in the SMC work because of its small contribution at
low energy. However,
this term is sizable in the $\Delta$-resonance region as can be seen
in Figure~\ref{ctil} which displays the $\acute{c}$ parameter along
with the dominant $c_0$ one.

\subsection{Pion absorption corrections\label{abso}}

\begin{figure}
\includegraphics[totalheight=5.5in,angle=-90]{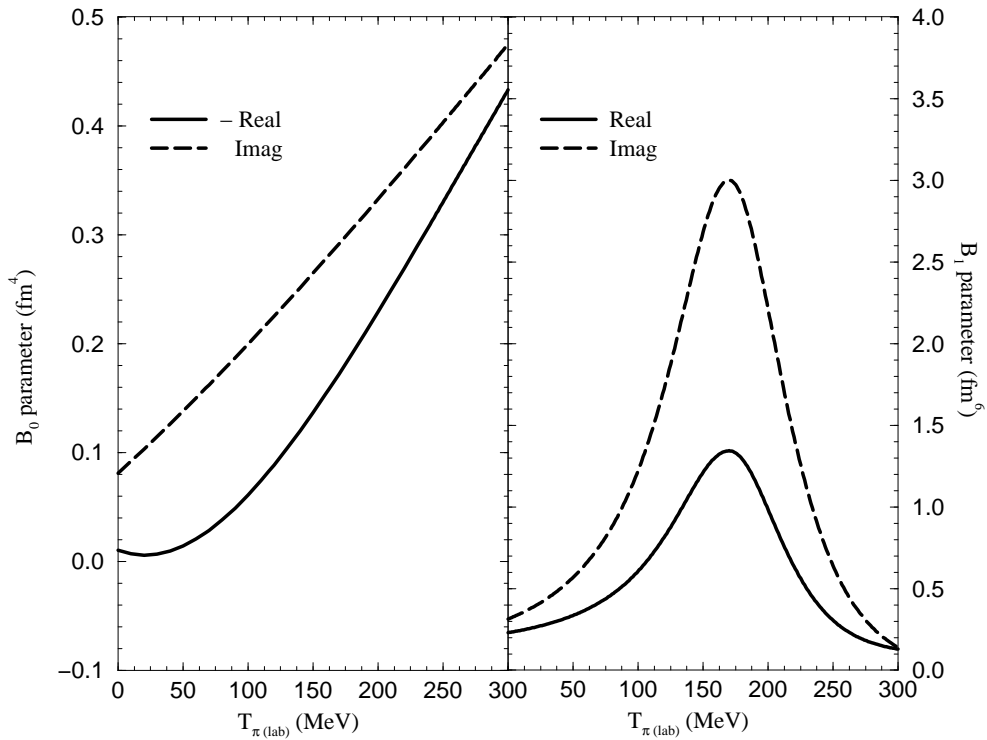}
\caption{\label{BC} The $B_0$ (left panel) and $C_0$ (right panel)
absorption parameters as functions of the pion kinetic
energy in the laboratory system $T_{\pi\;(\text{lab})}$.}
\end{figure}

Having included multiple scattering corrections, we are now in a
a place to discuss pion absorption processes that generate the
absorption terms in the optical potential.
There are two types of such terms: the first one arises from the fact that there
are many open inelastic channels in the $\pi$-nucleus interaction such as
nucleon knock-out. Accordingly, a portion of the incoming flux is absorbed
by these processes leading to an imaginary part in the potential. This
kind of absorption is naturally included in the impulse-approximation
form for the potential. The second type of absorption originates from
many-body mechanisms such as two-nucleon absorption where the pion is
scattered from one nucleon but then absorbed by another. This is in
fact the dominant many-body absorption mechanism and has to be
incorporated in the potential. The many-body absorption mechanisms are referred to
as ``true absorption'' to distinguish them from the inelastic (type one)
absorptions. Ironically, the $\Delta$-resonance formation that drives
strongly the elementary process $\pi N \rightarrow \pi N$, dampens it
in the nuclear medium through absorption channels.

The two-nucleon absorption contribution is derived from the
pion-two nucleon amplitude in analogous methodology to that of
Section~\ref{piN}, but in the pion-two nucleon c.m.
frame~\cite{erer66}. This introduces new scattering $s$- and $p$-wave
parameters $B_0$, $B_1$, $C_0$, and $C_1$ such as those in
Eq.~\ref{pinamp}. They can, as a matter of principle, be determined from the various
amplitudes of the pion-two nucleon systems including the reaction $\pi d
\rightarrow N N$. However, it is difficult to disentangle the
deuteron structure effects and thus we adopt the parameters as calculated
theoretically for nuclear matter by Chai and Riska~\cite{cr79}.

After establishing the pion-two nucleon amplitude, the expression is
transformed to the $\pi$-nucleus c.m. system and the angle transformation
is invoked to express the various kinematical quantities in the
$\pi$-nucleus c.m. frame. In turn, this introduces further kinematic
corrections such as those described in Section~\ref{kinc}.
The amplitude is then folded in the nucleus using the impulse
approximation. Higher order $p$-wave absorption terms are summed
leading to an absorption contribution to the Ericson-Ericson non-local
term (see Eq.~(\ref{eee})). Figure~\ref{BC} shows the $B_0$ and $C_0$ 
absorption parameters as a function of the pion kinetic
energy in the laboratory system. The absorption isospin
parameters $B_1$ and $C_1$ are very small and have been neglected
in our formalism.

\subsection{Pauli correction}

\begin{figure}
\includegraphics[totalheight=5.5in,angle=-90]{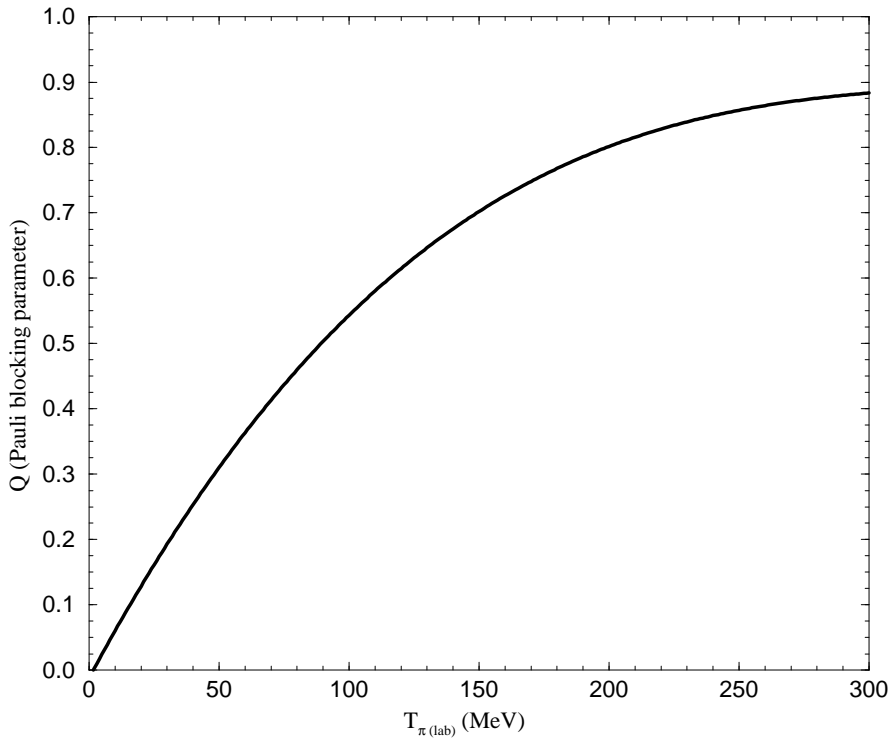}
\caption{\label{Q} The Pauli blocking parameter $Q$ as a function of the pion kinetic
energy in the laboratory system $T_{\pi\;(\text{lab})}$. This
parameter reduces the imaginary parts of the parameters $b_1$, $c_0$, and $c_1$ by a
factor of $Q$ to account for the reduction in bound-nucleon available
phase space.}
\end{figure}

Another alteration to the potential is the Pauli correction. Due to
the Pauli principle, the number of available final states for the
struck nucleon in the nuclear medium is reduced by Pauli blocking leading to
this kind of correction. The effect can be treated approximately by reducing
the imaginary parts of the parameters $b_1$, $c_0$, and $c_1$ by a
factor $Q$ that parameterizes the fraction of phase space available to the
struck nucleon~\cite{SMC2}. This  correction has already been
incorporated for the $b_0$ parameter in the calculation of the second order
correction. To be noted here that this effect
does not, to a large extent, influence the imaginary parts of the absorptions
parameters $B_0$, $B_1$, $C_0$, and $C_1$ as the nucleon gains a
large value of momentum when the pion is absorbed.

The $Q$ factor has been estimated for pions by Landau
and McMillan~\cite{lm73} and is shown for completeness in Figure~\ref{Q}. It is evident
that at low energy, $Q$ is vanishing as no sates are available for the
struck nucleon while it approaches the identity at higher energies as a large volume of
phase space becomes available to the struck nucleon.

\subsection{Coulomb corrections}

\begin{figure}
\includegraphics[totalheight=5.5in,angle=-90]{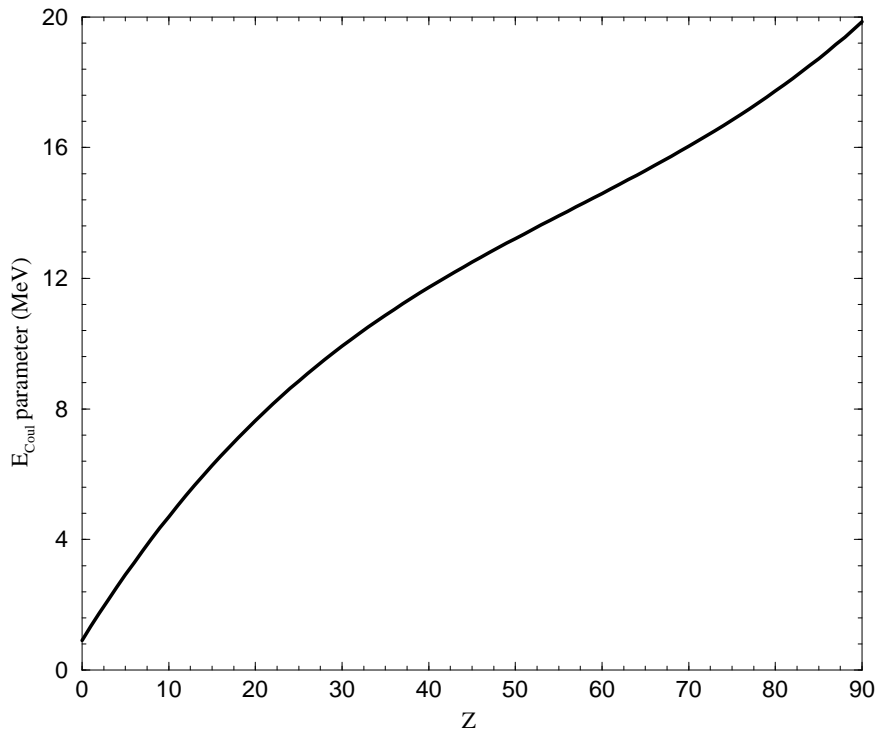}
\caption{\label{Coul} The Coulomb energy correction parameter
$E_{Coul}$ as a function of the nuclear proton number $Z$. This
correction shifts the value of the pion kinetic energy in the
laboratory system $T_{\pi\;(\text{lab})}$ at which the optical
potential is evaluated to account for the acceleration or deceleration
of the incoming charged pion, before it feels the short-range strong
interaction of the nucleus.}
\end{figure}

An additional correction is the Coulomb one,
stemming from the fact that the incoming charged pion (in
$\pi$-nucleus scattering) is accelerated or decelerated depending on
its charge, by the long-range Coulomb field of the nucleus before interacting
through the short-range strong interaction. This correction shifts the
value of kinetic energy at which the optical potential is evaluated
to account for the acceleration or deceleration according to $
T_{\pi\;(\text{lab})} \rightarrow T_{\pi\;(\text{lab})} - \varepsilon_\pi E_{Coul}$, where $E_{Coul}$ is the
value of the Coulomb field at the nuclear surface and
$\varepsilon_\pi$ is the pion charge. Figure~\ref{Coul}
displays the $E_{Coul}$ parameter as a function of
the proton number $Z$ of the nucleus. While this correction is very
small for
light nuclei and at low energies where the parameters vary slowly,
it plays a significant role for heavy nuclei and specially at the
$\Delta$-resonance region where the parameters change rapidly.

\section{Optical potential form}
\label{sec:form}

In implementing the above mentioned corrections, we arrive at a $\pi$-nucleus optical potential---applicable
from threshold up to the $\Delta$-resonance region---of the form:
\begin{eqnarray}
        2 \omega U &\!=\!&\!-4\pi \Big[ p_1 b(r) \!+\! p_2 B(r)\!-\!
        \mbox{\boldmath ${\nabla}$ \unboldmath} {\cal Q}(r) \cdot
        \mbox{\boldmath ${\nabla}$ \unboldmath} \\ \nonumber && \!-\!
        \; \frac {1}{4} p_1 u_1 {\nabla}^2 c(r)\!-\!  \frac {1}{4} p_2
        u_2 {\nabla}^2 C(r)\!+\!  p_1 y_1 \widetilde{K}(r) \Big]\;,
\end{eqnarray} 
where
\begin{eqnarray}
   b(r) &=& \bar{b}_0\rho(r)-\epsilon_\pi {b_1} {\delta}\rho(r) \;, \\
   B(r) &=& {B_0}\rho^2(r)-\epsilon_\pi {B_1}\rho(r)\delta\rho(r) \;,
   \\ c(r) &=& c_0\rho(r)-\epsilon_\pi {c_1}{\delta}\rho(r)\;, \\ C(r)
   &=& C_0\rho^2(r)-\epsilon_\pi{C_1}\rho(r){\delta}\rho(r)\;, \\ {\cal
   Q}(r)
   &=& \frac{ L(r) }{1 + \frac {4 \pi}{3} \lambda L(r)} + p_1 x_1
   \acute{c} \rho(r)\;, \\ L(r) &=& p_1 x_1 c(r) + p_2 x_2 C(r) \;, \\
   \widetilde{K}(r) &=& \frac {3}{5} \left({\frac{3
   \pi^2}{2}}\right)^{2/3} c_0\rho^{5/3}(r)\;,
\end{eqnarray}
and with
\begin{eqnarray}
   \bar{b}_0 &=& {b_0} - p_1 \frac {A-1}{A} (b^2_0 + 2 b^2_1) \:I \;, \\
   \acute{c} &=& p_1 x_1 \frac {1}{3} k^2_o (c^2_0 + 2 c^2_1) \:I \;.
\end{eqnarray}
In the above expressions, the set \{$p_1, u_1, x_1,$ and $y_1$\}
represents various kinematic factors in the effective $\pi-N$ system
(pion-nucleon mechanisms) that are determined from the
set of equations:
\begin{subequations}
\begin{eqnarray}
p_1 &=& \L(\frac{E_N + \omega} {E_A + \omega}\R) \L(\frac{E_A}{E_N}\R)\;,\\
u_1 &=& 2\: (D_1 + D_1^2)\;,\\
x_1 &=& (1 + D_1)^2\;,\\
y_1 &=& D_1^2\;.
\end{eqnarray}
\end{subequations}
Here
\begin{subequations}
\begin{eqnarray}
D_1 &=& \frac{F_1}{E_N + \omega}\;,\\
F_1 &=& \gamma_1 \L( \frac{\gamma_1}{\gamma_1 + 1} \vec{\beta}_1 \cdot \vec{k} - \omega \R)\;,\\
\gamma_1  &=& \frac{1}{\sqrt{1 - \beta_1^2}}\;,\\
\beta_1  &=& \frac{k}{E_N + \omega}  \L( 1 - \frac{1}{A} \R)\;,\\
\vec{\beta}_1 \cdot \vec{k}  &=& \beta_1 k\;.
\end{eqnarray}
\end{subequations}
In these expressions, $A$ is the atomic number, $\omega$ and $k$ are the pion energy and
momentum, while $E_A$ is the nucleus energy and $E_N$
is the nucleus energy per nucleon. These kinematic quantities
are in the $\pi$-nucleus c.m. frame. Note the appearance of the Lorentz
transformation parameters $\beta_1$ and $\gamma_1$ in these
expressions. This is a result of the angle transformation
described in Section~\ref{kinc}.

The set \{$p_2, u_2,$ and $x_2$\}
represents kinematic factors in the $\pi\!-\!2N$
system (pion-two nucleon mechanisms): 
\begin{subequations}
\begin{eqnarray}
p_2 &=& \L(\frac{2 E_N + \omega}{E_A + \omega }\R) \L(\frac{E_A}{2 E_N}\R)\;,\\
u_2 &=& 2 \:(D_2 + D_2^2)\;,\\
x_2 &=& (1 + D_2)^2\;.
\end{eqnarray}
\end{subequations}
where 
\begin{subequations}
\begin{eqnarray}
D_2 &=& \frac{F_2}{2 E_N + \omega}\;,\\
F_2 &=& \gamma_2 \L( \frac{\gamma_2}{\gamma_2 + 1} \vec{\beta}_2\cdot \vec{k}- \omega_2 \R)\;,\\
\gamma_2  &=& \frac{1}{\sqrt{1 - \beta_2^2}}\;,\\
\beta_2  &=& \frac{k}{2 E_N + \omega}  \L( 1 - \frac{2}{A} \R)\;,\\
\vec{\beta}_2\cdot \vec{k}  &=& \beta_2 k\;.
\end{eqnarray}
\end{subequations}
Note the appearance of the factor ``2'' in many terms in these pion-two nucleon
quantities as opposed to the factor ``1'' in the corresponding quantities in the
pion-nucleon system.

The set of parameters \{$b_0,
b_1, c_0,$ and $c_1$\} originates from the $\pi N\!\rightarrow\!\pi N$
elementary amplitudes while all other parameters, excluding kinematic
factors, have their origin in the second and higher order
corrections to the optical potential. Nuclear effects enter the
optical potential through the nuclear density $\rho(r)$, and through
the neutron-proton density difference (isovector density)
$\delta\rho(r)$. These densities were calculated using a
mean-field approximation to the quantum hadrodynamics model (QHD) of Walecka~\cite{SW86}. Finally, $k_o$ is the
pion laboratory momentum, $\lambda$ is the
Ericson-Ericson effect parameter, and $I$ is the $1/r_{correlation}$ function (Section~\ref{muls}). The $B$ and $C$
parameters arise from true pion absorption (Section~\ref{abso}).

\section{Results and Discussion}
\label{sec:results}

\begin{figure}
\includegraphics[totalheight=5.5in,angle=-90]{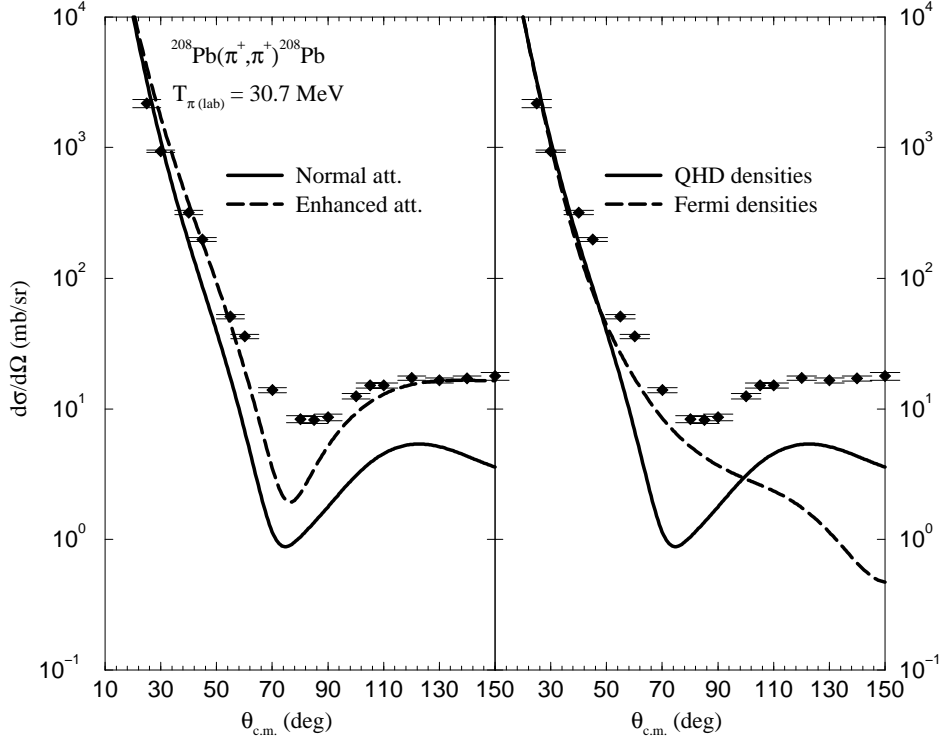}
\caption{\label{Pb} The differential cross section of $\pi^+$ elastic
scattering from ${}^{208}$Pb as a function of the scattering
angle in the c.m. frame at pion laboratory kinetic energy of
30.7~MeV. Experimental data are obtained from
Ref.~\cite{preedom81}. The left panel displays the results with normal
(solid line) and enhanced (dashed line) $s$-wave
attractions. The right panel shows the cross section with Walecka
model (QHD)
calculations for the nuclear desnities (solid line) and with a two-parameter
Fermi density (dashed line). The figure illustrates the effects of missing low-energy
$s$-wave strength as well as the uncertainty arising from different
calculations for the nuclear structure.}
\end{figure}

\begin{figure}
\includegraphics[totalheight=5.5in,angle=-90]{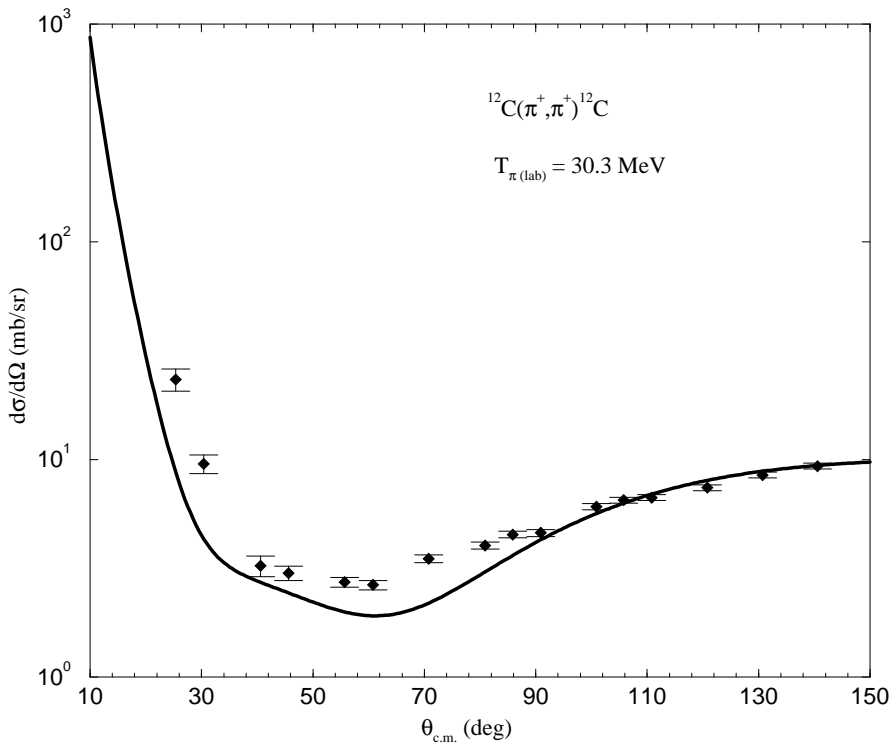}
\caption{\label{C} The differential cross section of $\pi^+$ elastic
scattering from ${}^{12}$C as a function of the scattering angle in
the c.m. frame at pion laboratory kinetic energy of
30.3~MeV. Experimental data are obtained from
Ref.~\cite{preedom81}. This figure provides an application of the
optical potential at low energy.}
\end{figure}

Having establsihed the optical potential, it is appropriate to
test it for a variety of nuclei and at various energies from threshold
up to the $\Delta$-resonance region. Figure~\ref{Pb} shows the differential
cross section of $\pi^+$ elastic scattering from ${}^{208}$Pb at pion
laboratory kinetic energy of 30.7~MeV and as a function of the
scattering angle in the c.m. frame. Experimental data are obtained
from Ref.~\cite{preedom81}. In the left panel of the figure, we
display the calculated cross section using our optical potential
(solid-line) and comapre it with the data. It is evident that the
potential lacks some strength as it underestimes the
data. This is in fact the well-known problem of missing $s$-wave attraction at low
energy~\cite{SMC1,SMC2,SMC3,SMC4}. Indeed, this missing strength
was one of the main reasons that caused the SMC group to refit the
parameters enhancing the $s$-wave attraction.  We
also show in the same panel the significantly improved result with an enhanced $s$-wave
attraction. The
physics behind this enhancement is not clear, but likely due
to some many-body effect not included in the optical
potential. The effect is specially pronounced for heavy nuclei such as
${}^{208}$Pb and does not appear to be as important for light ones as can be seen in
Figure~\ref{C} where we plot the differential
cross section of $\pi^+$ elastic scattering from ${}^{12}$C at pion
laboratory kinetic energy of 30.3~MeV. The experimental
data, taken from Ref.~\cite{preedom81}, are
well reproduced.

It is necessary here to stress one caution in comparing theoretical
 results with experimentla data. The derived optical potential depends
 on the matter and iso nuclear desnities. These densities are
 independent of the reactive content of the $\pi$-nucleus interaction but
 depends solely on the static properties of nuclei. They can be calculated
 with a variety of models with different sophistication. This in turn
 introduces an element of ambiguity in comparing to experiemntal data
 as we cannot determine whether any discripanices with the data are
 due to the optical potential or to the nuclear density
 calculations. In the right panel of Figure~\ref{Pb}, we exhibit two
 calculations, one using Walecka model (QHD)~\cite{SW86} while the
 other using the simple two-parameter Fermi densities~\cite{PRSZ95}. It is
 evident that there are significant differences at large angles
 attributed to differences in the large momentum components of the
 densities in these two models. In order to minimize, if not
 eliminate, the ambiguities arising from nuclear desnities, and in
 order to concentrate our invistigation on the reactive content of the
 $\pi$-nucleus interaction, we
 have used the sophisticated Walecka model for calculating the nuclear
 structure.

\begin{figure}
\includegraphics[totalheight=5.5in,angle=-90]{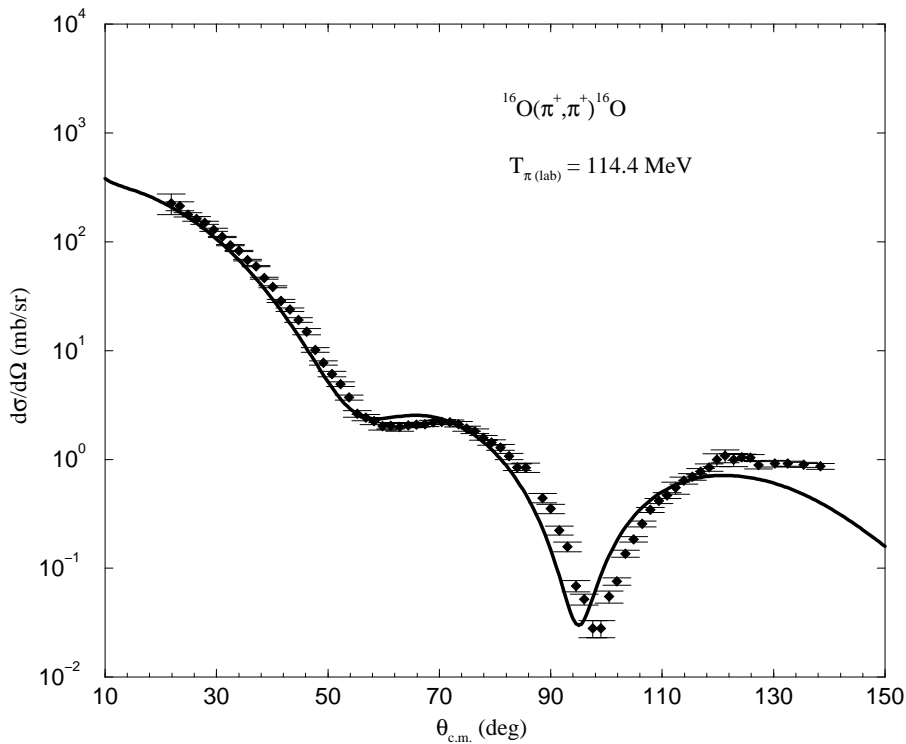}
\caption{\label{O1} The differential cross section of $\pi^+$ elastic
scattering from ${}^{16}$O as a function of the scattering angle in
the c.m. frame at pion laboratory kinetic energy of
114.4~MeV. Experimental data are obtained from
Ref.~\cite{Alba80}. The figure provides an application of the
optical potential at intermediate energy.}
\end{figure}

\begin{figure}
\includegraphics[totalheight=5.5in,angle=-90]{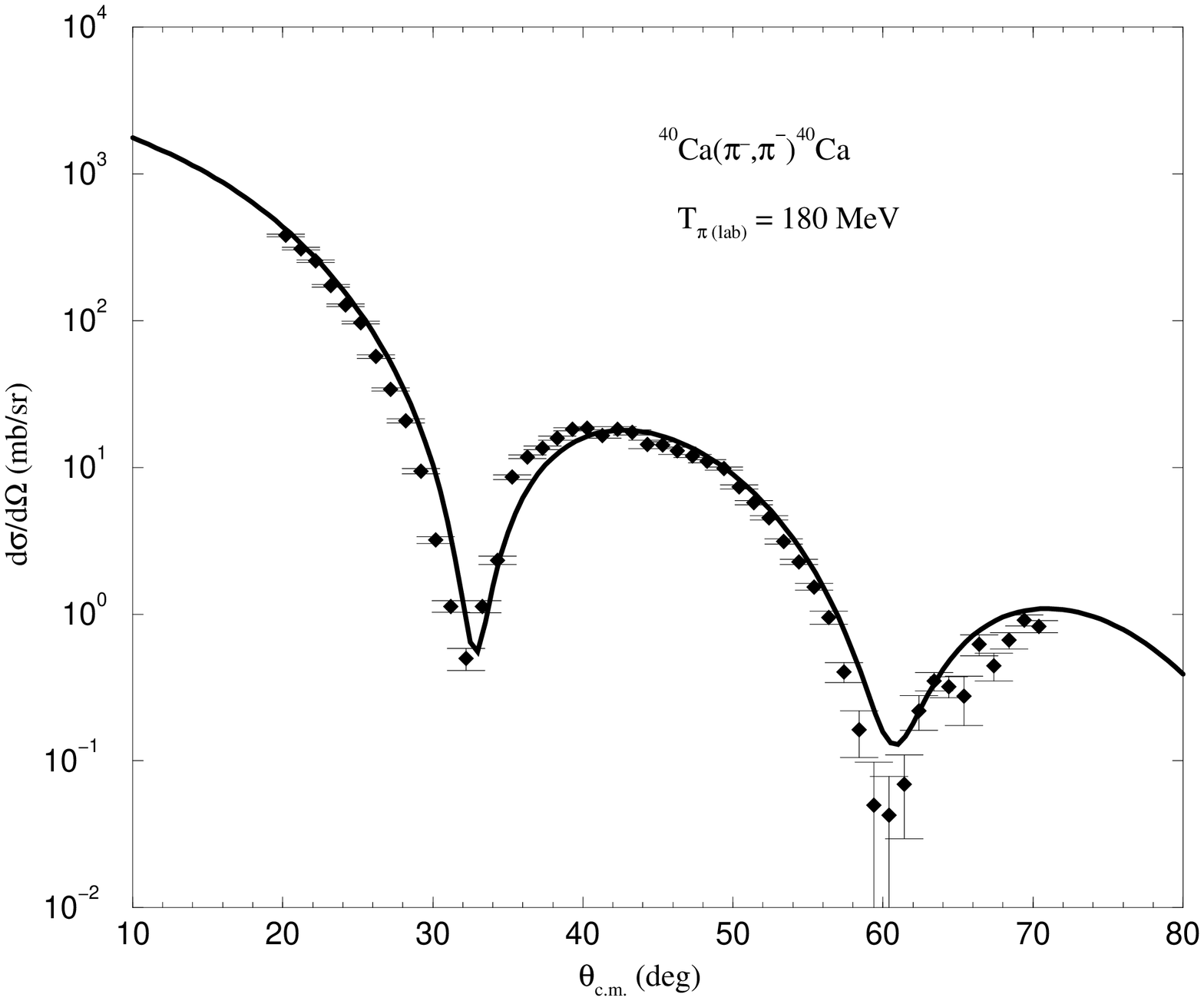}
\caption{\label{Ca} The differential cross section of $\pi^-$
elastic scattering from ${}^{40}$Ca as a function of the scattering
angle in the c.m. frame at pion laboratory kinetic energy of
180~MeV. Experimental data are obtained from Ref.~\cite{b84}. The
figure provides an application of the optical potential close to the
$\Delta$-resonance peak.}
\end{figure}

\begin{figure}
\includegraphics[totalheight=5.5in,angle=-90]{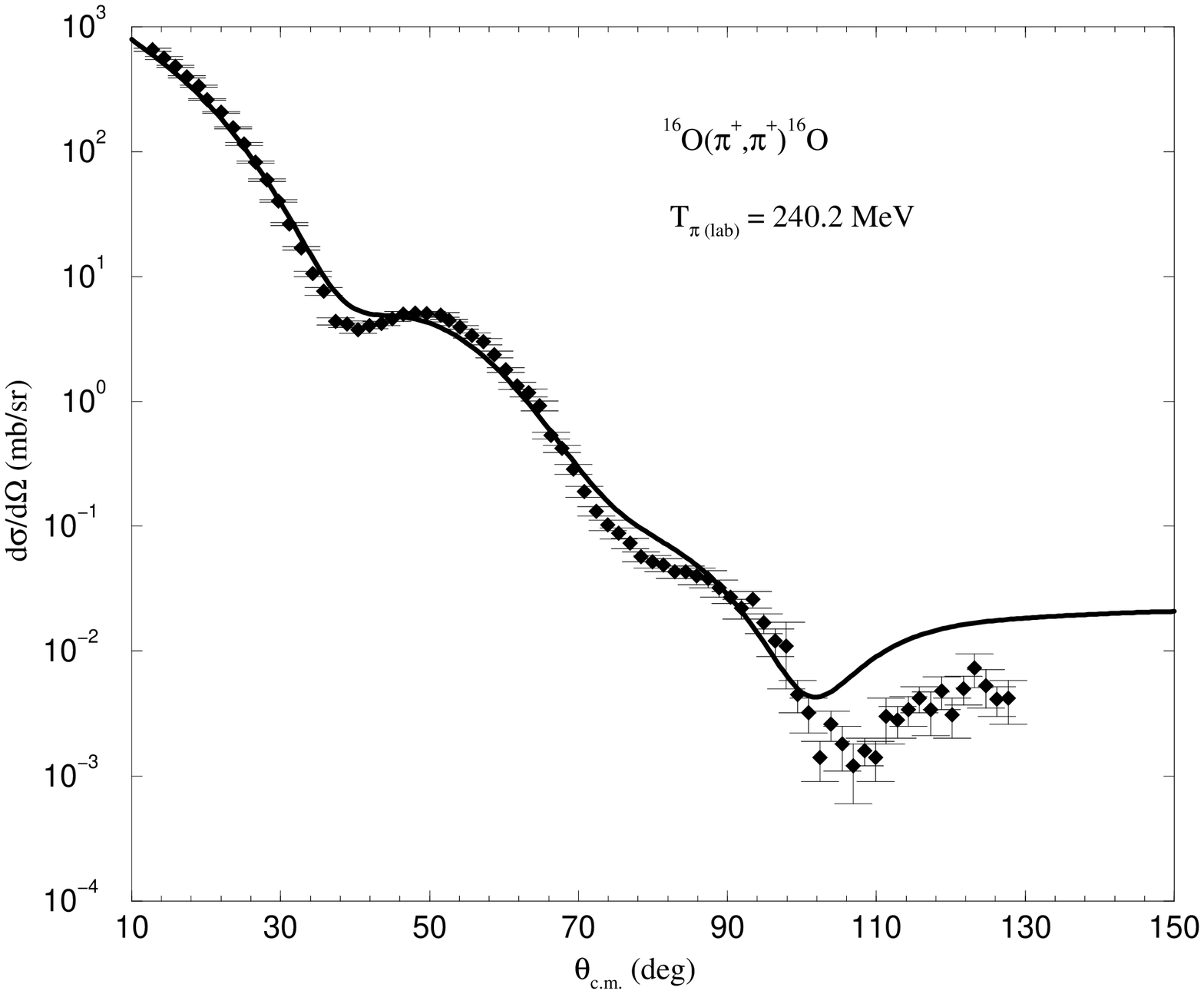}
\caption{\label{O2} The differential cross section of $\pi^+$ elastic
scattering from ${}^{16}$O as a function of the scattering angle in
the c.m. frame at pion laboratory kinetic energy of
240.2~MeV. Experimental data are obtained from
Ref.~\cite{Alba80}. The figure provides an application of the
optical potential towards the end of the $\Delta$-resonance region.}
\end{figure}

Having provided an application of our optical potential at low
energy, we proceed with comparisons at higher energies for a
variety of nuclei. In Figure~\ref{O1} we show a comparison for
${}^{16}$O at at pion
laboratory kinetic energy of 114.4~MeV, with experimental data from
Ref.~\cite{Alba80}. In Figure~\ref{Ca}, we exhibit
the results for ${}^{40}$Ca at laboratory kinetic energy of 180~MeV
(close to the $\Delta$-resonance peak) with data obtained from Ref.~\cite{b84}. Finally in
Figure~\ref{O2}, we show a comparison for ${}^{16}$O at 240.2~MeV,
that is towards the end of the $\Delta$-resonance region. The data are taken from
Ref.~\cite{Alba80}. It is evident that the experimental data are well produced by the inherent dynamics of the
potential. One should stress here that this agreement is generated
using a physics-based potential rather than a phenomenological one
with free parameters to be chosen. All parameters in our treatment have
their origin in the physics behind the scattering process.
One in principle can modify the parameters in order to fit any specific nucleus,
but we prefer this fundamental approach with its innate unity and
clear conception. 

\section{Conclusions}
\label{sec:concl}

In conclusion, based on an earlier seminal treatment of the low-energy
$\pi$-nucleus optical potential, we have derived a potential form
applicable from threshold up to the $\Delta$-resonance region. We have
done so by deriving the impulse approximation expression and then adding to
it kinematical and physical corrections. The kinematical ones arise
from transforming the impulse approximation expression from the $\pi-N$
c.m. frame to the $\pi$-nucleus c.m. system, while the physical ones
stem mostly from the many-body nature of the $\pi$-nucleus
interaction. By taking advantage of experimental advances in our
knowledge of the $\pi-N$ process, we have updated earlier treatments
with parameters calculated from state-of-the-art experiemntal
measurements. In this manner, we provide a physics-based optical
potential that can be used in diverse studies involving the
$\pi$-nucleus interaction.

\begin{acknowledgments}
 I am very grateful to Professor J. A. Carr for the
many informative discussions we have had on this subject. Furthermore,
I would like to thank Professor G. C. Hillhouse for his advice on
Coulomb distortions as well as Professor R. J. Peterson for supplying
me with some of the experimental data.  This work was supported by the
Japan Society for the Promotion of Science and the United States
National Science Foundation under award number 0002714.
\end{acknowledgments}

\bibliography{optical}
\end{document}